\documentclass[iop]{emulateapj}
\usepackage{epstopdf}

\newcommand\pasa{{PASA}}

\def\etal{{et~al.}\ }

\def\kpc{{\rm \; kpc}}

\def\msun{{M$_\odot$}}

\shorttitle{Hot Gas Condensation}

\begin{document}
\DeclareGraphicsExtensions{.pdf,.gif,.jpg,.eps}

\title{Gas Condensation in the Galactic Halo} 

\author{M.~Ryan~Joung\altaffilmark{1,2}, Greg~L.~Bryan\altaffilmark{1}, and Mary~E.~Putman\altaffilmark{1}}
\altaffiltext{1}{Department of Astronomy, Columbia University, 550 West 120th Street, New York, NY~10027; moo@astro.columbia.edu}
\altaffiltext{2}{Department of Astrophysics, American Museum of Natural History, 79th Street at Central Park West, New York, NY~10024}


\begin{abstract}
Using adaptive mesh refinement (AMR) hydrodynamic simulations of vertically stratified hot halo gas, we examine the conditions under which clouds can form and condense out of the hot halo medium to potentially fuel star formation in the gaseous disk.  We find that halo clouds do not develop from linear isobaric perturbations.  This is a regime where the cooling time is longer than the Brunt-V\"ais\"al\"a time, confirming previous linear analysis. 
We extend the analysis into the nonlinear regime by considering mildly or strongly nonlinear perturbations with overdensities up to 100, also varying the initial height, the cloud size, and the metallicity of the gas.  Here, the result depends on the ratio of cooling time to the time required to accelerate the cloud to the sound speed (similar to the dynamical time).  If the ratio exceeds a critical value near unity, the cloud is accelerated without further cooling and gets disrupted by Kelvin-Helmholtz and/or Rayleigh-Taylor instabilities.  If it is less than the critical value, the cloud cools and condenses before disruption.  
Accreting gas with overdensities of 10--20 is expected to be marginally unstable;  the cooling fraction will depend on the metallicity, the size of the incoming cloud, and the distance to the galaxy.  Locally enhanced overdensities within cold streams have a higher likelihood of cooling out.  
Our results have implications on the evolution of clouds seeded by cold accretion that are barely resolved in current cosmological hydrodynamic simulations and absorption line systems detected in galaxy halos.  
\end{abstract}



\section{Introduction}
The Milky Way and other similar-sized star-forming disk galaxies will exhaust their cold gas reservoirs within a few Gyrs unless it is replenished in some way (e.g, Chiappini et al. 2001; Sommer-Larsen et al. 2003).  One possible source of this fuel is the direct accretion of cold gas from the halo (e.g., Wakker \& van Woerden 1997; Thilker et al 2004; Oosterloo et al. 2007); however, this level of fueling does not appear to be sufficient to balance the star formation rate (e.g., Putman 2006; Sancisi et al 2008; Grcevich \& Putman 2009).

An additional possible source of fuel is gas accretion from an extended hot ($T \sim 10^6$ K) halo that is thought to surround disk galaxies (e.g., White \& Frenk 1991; Fukugita \& Peebles 2006; Bregman 2007).  Although the size of this reservoir is not well-constrained, the observational indications for its existence are many.  These include the structure of many high velocity clouds (HVCs; Bruens et al. 2000; Peek et al. 2007; Putman et al. 2011), O{\small~VI} absorption indicating the HVCs are moving through a hot halo medium (Sembach et al. 2003), the likelihood that the Magellanic Stream shows signs of ram-pressure effects (Weiner \& Williams 1996; Putman \etal 2003; Bland-Hawthorn \etal 2007), as well as the lack of gas in any dSph galaxy within $\sim$300 kpc of the Milky Way (Grcevich \& Putman 2009).  Observational constraints on the temperature and density of the hot gas also come from O{\small~VII} absorption lines in Chandra and XMM observations of distant quasars and halo objects (Wang et al. 2005; Bregman 2007).  The exact location of this O{\small~VII}-absorbing hot gas remains under some debate, but based on the observed line-strength vs. Galactic latitude, the majority of the absorption signal is construed to lie within $\sim$10 kpc of the plane (Rasmussen \etal 2003; Wang \etal 2005; Collins \etal 2005).  On the other hand, theoretical models, and some of the observational results mentioned above, indicate a very diffuse component of the hot gaseous halo extending out beyond 100 kpc (e.g., Maller \& Bullock 2004; Kaufmann \etal 2006; Fukugita \& Peebles 2006; Sommer-Larsen 2006).  The potential mass reservoir in this component may therefore be substantial (but see Anderson \& Bregman 2010).

Whether or not this hot halo gas can cool and accrete on to the disk is unclear.  One possibility is that the gas condenses out directly due to a thermal instability (e.g., Maller \& Bullock 2004).  This idea was examined in numerical simulations using the SPH method and precisely such condensation was found (Kaufmann \etal 2006, 2009; Sommer-Larsen 2006) with properties similar to those of the observed population of HVCs (Peek \etal 2008).  This is appealing as it explains both the nature and source of the HVCs as well as the fuel for ongoing star formation in the disk of the Milky Way.

However, this simple condensation picture has recently been challenged. A linear analysis of the thermal instability operating in a stable, stratified atmosphere indicates that the gas is not likely to be unstable unless the entropy profile is extremely flat (Binney \etal 2009 and references therein).  This is unlike the case in an homogeneous medium (e.g. Field 1965; Burkert \& Lin 2000).  The reason is that oscillations driven by the buoyancy of the thermally unstable gas will disrupt the gas parcel before it can cool.   A similar situation has been found to exist in clusters of galaxies (e.g. Malagoli et al. 1987; Balbus \& Soker 1989).  If the entropy profile is flat, or very close to flat, the oscillation timescale can be longer than the cooling time and instabilities can occur; however, it remains unclear if such profiles actually occur in nature, or if they can persist for a long period of time.

A possible way around this objection is to have the perturbations in the gas be nonlinear, such as might be found in inflowing filamentary gas which then cools to form clouds, as suggested in Kere\v{s} \& Hernquist (2009).  This is appealing as it links the HVCs and gas replenishment directly to the relatively cool ($T \sim 10^5$ K), filamentary flows found in galaxy formation simulations (Katz \etal 2003; Kere\v{s} \etal 2005; Ocvirk \etal 2008; Brooks \etal 2009; Kere\v{s} \etal 2009).  Nonlinear perturbations could also result from the leftovers of satellite accretion (e.g., Bland-Hawthorn et al. 2007; Grcevich \& Putman 2009).

In this paper, we numerically investigate the cooling of instabilities in a stratified atmosphere.  We use parameters appropriate for the Milky Way hot halo, and investigate both linear and nonlinear instabilities in order to understand when gas can cool and condense out of the hot gas.

\section{Method}

We perform simulations with Enzo, an Eulerian hydrodynamics code with adaptive mesh refinement (AMR) capability (Bryan \& Norman 1997; Norman \& Bryan 1999; Bryan 1999; O'Shea et al. 2004).  It solves the Euler equations using the piecewise-parabolic method (PPM; Colella \& Woodward 1984) to handle compressible flows with shocks.  We choose to use PPM rather than the Zeus hydro solver, which is also included in Enzo, primarily because the former is known to resolve instabilities better (e.g., Agertz et al. 2007).

We set up a stratified isothermal gas at 2.0 $\times$ 10$^6$ K initially in hydrostatic equilibrium.  This value represents our best guess for the temperature of the observed hot halo gas (e.g., Henley et al. 2010).  The computation box contains a volume of (1 kpc)$^2$ $\times$ (20 kpc), elongated in the vertical direction, in order to follow the evolution of clouds as they fall towards the midplane (located in the middle of our volume).  

The simulations include metallicity-dependent radiative cooling (Cen et al. 1995) extended down to 10 K (Dalgarno \& McCray 1972), while evolving a chemical gas-phase reaction network of nine species (Abel et al. 1997).  A diffuse form of photoelectric heating (Wolfire et al. 1995, 2003) that scales weakly with gas density (Joung et al. 2009) is included for gas with $T \le 10^4$ K, and so is applied only when the cloud cools sufficiently.

In most of our models, we place up to 3 levels of refinement on top of a base grid of 32$^2$ $\times$ 640 cells, resulting in an effective resolution of 256$^2$ $\times$ 5120 cells, although we limit refinement only to the top half of the box ($z \ge 0$).  Our refinement criteria are based on gas density and cooling time.  Specifically, the refinement process ensures that the mass contained in a given cell is less than 4.9 $M_{\odot}$ (in the fiducial model) and that the cooling time in a cell is always shorter than the sound crossing time of that cell.  This criterion ensures that the cooling time is well resolved.  For simplicity, star formation and energy input from stars are not included in our models.  We also do not include heat conduction; see section \ref{sec:caveats} for a discussion of this point.

\begin{figure}
\epsscale{0.8}
\hspace{0mm} \plotone{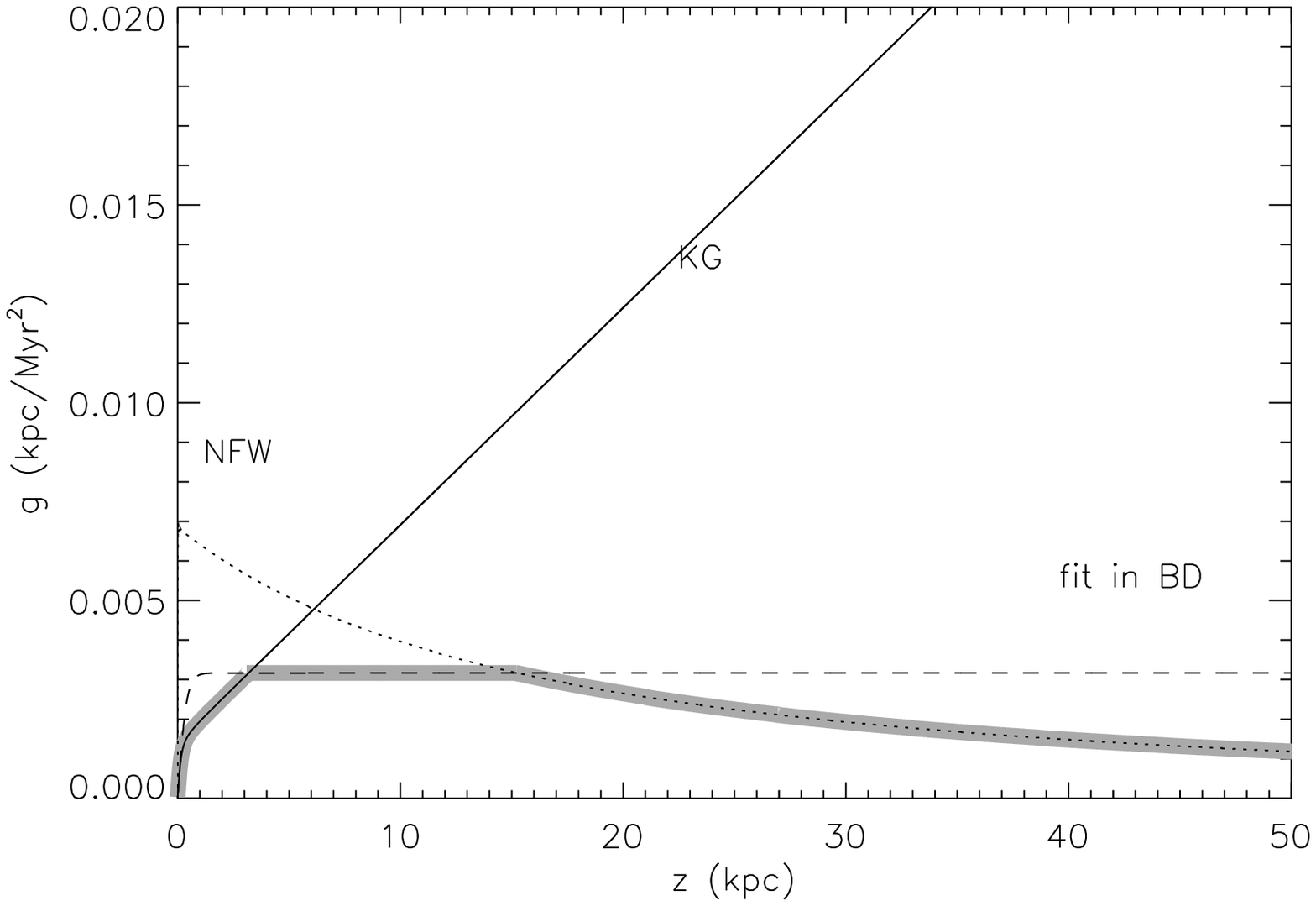}
\caption{Vertical gravitational acceleration adopted in our models.  The thick light grey curve represents our choice of $g(z)$ at different ranges of heights when each analytic expression is thought to be valid;  see equation (\ref{gz}).  In the figure, NFW indicates the (radial) acceleration from Navarro et al (1997), KG is the vertical acceleration from Kuijken \& Gilmore (1989), appropriate for $z \lesssim 3$ kpc, and the line marked BD is from Benjamin \& Danly (1997).
\label{gofz}}
\end{figure}

The gravitational influence of the disk and dark matter halo are included through a static gravitational term, which is strictly vertical.  The variation of the gravitational acceleration as a function of height, $g(z)$, is important as it determines both the initial density profile -- assumed to be in hydrostatic equilibrium -- and the Brunt-V\"ais\"al\"a (B-V) time, i.e. the period of oscillations due to buoyancy (defined below).  Kuijken \& Gilmore (1989) derived their estimate of $g(z)$ based on a sample of K dwarfs towards the South Galactic pole, which is applicable to $|z| \lesssim 3$ kpc in the solar neighborhood.  This is employed in our models for $|z| \le 3.06$ kpc.  For the layer immediately above it (3.06 kpc $\le |z| \le$ 15.2 kpc), we adopt an expression for the gravitational acceleration in Benjamin \& Danly (1997) used to determine the distances to falling interstellar clouds in the Galactic halo, which in turn was taken from Wolfire et al (1995).  The value of $g(z)$ within this given range of $z$ is nearly constant.  Finally, in the uppermost layer, the NFW profile (Navarro et al. 1997) for a halo with a mass of 2 $\times$ 10$^{12}$ $M_{\odot}$ and a concentration parameter, $c = 12$, is used, where applicable.  Hence,
\begin{equation}
\label{gz}
g(z) = \left\{
\begin{array}{ll}
 a_1 \frac{z}{\sqrt{z^2+z_0^2}} + a_2 z & |z| \le 3.06 \kpc  \\
 a_3 \tanh \left( \frac{z}{0.4 \kpc} \right) &   3.06 \kpc \le |z| \le 15.2 \kpc \\
 \frac{a_4 R_s^2}{z^2} \, \frac{ \ln (1 + z/R_s) - z/(R_s + z) }{ \ln (1 + c) - c/(1 + c) } &  |z| > 15.2 \kpc 
\end{array} \right.
\end{equation}
Here, $a_1 = 1.42 \times 10^{-3}$ kpc/Myr$^2$, $a_2 = 5.49 \times 10^{-4}$ Myr$^{-2}$, $z_0 = 0.18$ kpc, $a_3 = 9.5 \times 10^{-9}$ kpc/Myr$^2$, $a_4 = 6.97 \times 10^{-9}$ kpc/Myr$^2$, and $R_s = 20$ kpc.  The three curves as well as our choice of $g(z)$ are shown in Figure~\ref{gofz}.

\begin{figure}
\hspace{32mm} \includegraphics[scale=0.53,angle=90]{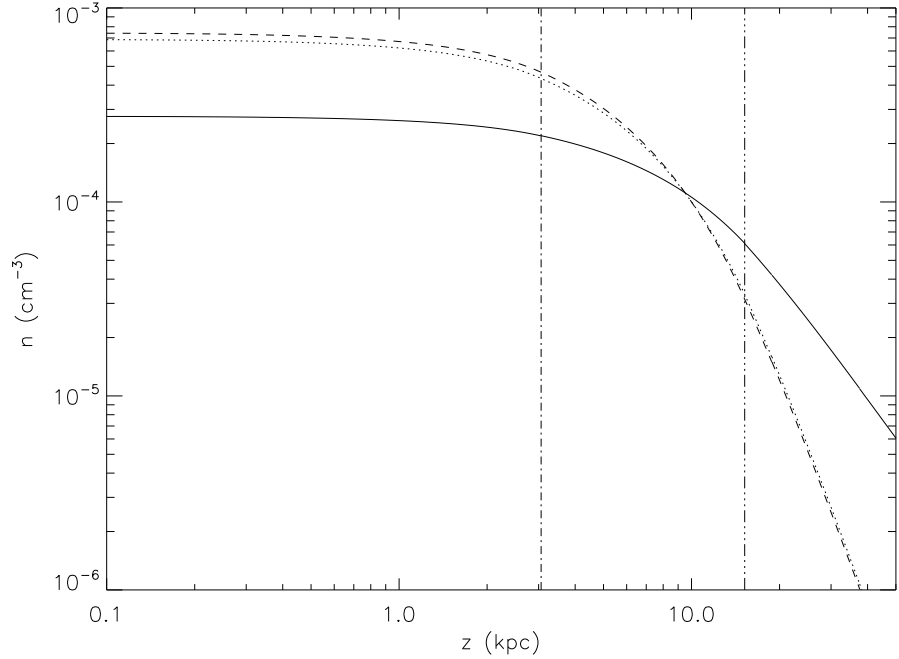}
\caption{Initial density profiles in hydrostatic equilibrium with the gravitational field in equation (\ref{gz}).  The density is normalized to $10^{-4}$ cm$^{-3}$ at $z = 10$ kpc.  The vertical lines at $z = 3.06$ and $15.2$ kpc indicate where the analytic expression for $g(z)$ changes.  The solid curve is the density profile for $T_{\rm halo} = 2 \times 10^6$ K and $\mu = 0.61$, as adopted in our models.  The dashed curve shows how the profile would change if we adopted a lower (but still consistent with observational constraints) value for $T_{\rm halo}$ of $1 \times 10^6$ K.  For the dotted curve, $T_{\rm halo} = 2 \times 10^6$ K is again assumed but $\mu = 1.2$.
\label{rhoz}}
\end{figure}

Figure~\ref{rhoz} shows the initial density profile in hydrostatic equilibrium with the gravitational field.  The overall density normalization is not constrained by hydrostatic equilibrium; the density is normalized so that $n = 10^{-4}$ cm$^{-3}$ at $z = 10$ kpc (Hsu et al. 2011; Gaensler et al. 2008; Rasmussen et al. 2009; Grcevich \& Putman 2009; McCammon et al. 2002; Sembach et al. 2003). 

Our model of an initially isothermal, hot gas distribution is obviously an over-simplification, since we know that near the midplane the mean gas temperature decreases significantly and the mean density is significantly higher.  However, this simplified model suffices for our purpose because we are interested in the early evolution of density perturbations, while they are well above the disk.  Note that including a cooler (e.g., $T \sim 10^4$ K), denser midplane layer would have introduced an additional difficulty at the interface between the hot and the cool layers, as the transition will be smoothed over a few cells, and the cooling rate will peak in those interface cells, which is unphysical.  Ideally, to set up a realistic ISM, mechanical (stellar winds and supernova explosions; see, e.g., Joung \& Mac Low 2006) and radiative feedback from star formation should be included, but this is beyond the scope of the present study.

Periodic boundary conditions are used on the side surfaces perpendicular to the Galactic plane, while outflow boundary conditions are used on the upper and lower surfaces.  We found that the outflow boundary conditions (more appropriately called 'zero gradient boundary conditions') do not lead to a static background medium, even if we set up the "active" region of the box ($|z| \le 10$ kpc in our fiducial model) in exact hydrostatic balance.  Instead, the background medium gradually developed inflow motions toward the midplane ($z = 0$), starting from the top and bottom boundaries, and after a sufficiently long period ($>100$ Myr), gas initially located at those boundaries moved significantly closer to the midplane.  The same behavior was seen even after radiative cooling was turned off.  This is because the outflow boundary conditions, which simply copy the variables in the outermost cell in the active region to the boundary cells, neglect the small yet important vertical pressure gradient caused by the density drop-off immediately outside of the upper and lower boundaries.  Note that previous models that used outflow boundary conditions in the vertical direction may have suffered from the same problem, although because of the global catastrophic collapse due to cooling on shorter timescale, the effect was hidden (see, e.g., de Avillez 2000; Joung \& Mac Low 2006).  As a simple remedy for the problem, we set the densities in the boundary cells not at the value in the outermost cell but at a value slightly lower than it by extending the density gradient by one more cell.  Doing that was enough to stabilize the background medium.

We introduce a density perturbation with a spherical shape (a ``cloud") that is overdense with respect to the background medium by a prescribed value, which is an input parameter of the models.   Note that the density inside the cloud is not uniform but is a function of height, as the background density decreases in the vertical direction.  The temperature is reduced by the same factor so that the cloud is initially in pressure equilibrium with the surrounding medium.

The initial height of the cloud in our fiducial model is 8 kpc, with a cloud radius of 0.25 kpc, and we use overdensities relative to the mean density of $\delta = \delta \rho/\rho =$ 1.01, 1.1, 2, 4, 7, 10 and 30 (see Table~\ref{tbl_models} for parameters for all models).  In addition, we ran some additional models with smaller and larger initial heights, ranging from 2 kpc to 32 kpc.  In runs with our largest heights (16 and 32 kpc), we increased both the clouds radius and the box size to twice and four times the fiducial size in every dimension, respectively, in order to also examine the impact of cloud size.    The diameter of the cloud is 1/2 of the width of the computation box, so the cloud size and its mass scale with the box size.
The largest height, 32 kpc, was chosen to be representative of distances where cold gaseous streams seed dense clouds aided by hydrodynamic instabilities in the scenario advocated by Kere\v{s} \& Hernquist (2009).  

Most simulations were run twice, once with primordial abundances (i.e. only H and He cooling), and once with a metallicity of 0.3 solar.  We ran additional models with $Z=0$, 0.1 and 0.5 $Z_{\odot}$ to study the effect of varying metallicity.  The metallicity of the gas is assumed to be uniform throughout the box.

\section{Results}

The simulations we run are described in Table~\ref{tbl_models}, where the naming convention encodes both the height and overdensity of the cloud.  For example, {\it h4o8} indicates a model with a cloud starting at 4 kpc with an overdensity of 8 relative to the background density at that height.

\subsection{Linear regime}

We begin by examining the results of the model with the smallest amplitude in density and temperature (only 1\% of the background values) and compare them with the linear predictions made by Binney et al. (2009).  They predicted that for this background entropy profile, and for perturbation amplitudes that are small and approximately isotropic, the perturbation should oscillate on the B-V time:
\begin{equation}
t_{\rm BV} = 2 \pi \left[ \frac{g}{T} \left( \frac{dT}{dz} + \frac{\gamma - 1}{\gamma} \frac{g\mu m_p}{k_B} \right) \right]^{-1/2},
\label{eq:BV_time}
\end{equation}
where $\mu$ is the mean particle mass and $\gamma = 5/3$ is the ratio of specific heats.  In our fiducial model with an initial cloud height of 8 kpc, the cooling time is longer than the B-V time by a factor of $\sim$10, and so the cloud (with only 1\% density enhancement) is expected to simply oscillate up and down. Half of the time, it is cooler than the surrounding gas and half of the time it is hotter, so, to first order there should be no net cooling, and the clump should not be unstable in the sense of Field (1965).   For the cooling time, we adopt:
\begin{equation}
t_{\rm cool} = \frac{3kT}{2n \Lambda(T)}
\label{eq:cool_time}
\end{equation}
where $\Lambda(T)$ is the cooling rate at temperature $T$.  These timescales are shown in Figure~\ref{timescale}, both as a function of clump overdensity and also initial clump height.

The time sequence in Figure \ref{oscillate} shows the $z$-component of the velocity, and confirms  that the cloud indeed oscillates vertically around the equilibrium position.  The grey dotted curve represents a simple sinusoidal wave with a period of 527.4 Myr, as predicted by linear analysis (eq. \ref{eq:BV_time}).  The dashed curve represents the mean vertical velocity of the cloud in the model without radiative cooling, while the solid curve shows the same quantity in the model with cooling.  The exact shape of the cloud is not retained since inside the initially uniform sphere various modes having different wavelengths interact with one another and distort the waveform.  For this reason, we plot the mean $z$-component of the velocity of the oscillation (the area under the $v_z$ vs. $x$ curve, divided by the wavelength, here assumed to be the initial diameter of the cloud\footnote{Only the cylindrical volume within $\Delta z \le 1.5$ kpc of the vertical line passing through the center of the cloud ($\Delta r \le 15.6$ pc) was used for this analysis.}).  

After an initial transient period ($t \lesssim 100$ Myr), the mean amplitude of the $z$-velocity displays a simple sinusoidal shape with a period of 540 Myr, close to the estimate from the linear analysis, 527.4 Myr.  The first half period of the cycle, when the cloud is on average below the initial position, is slightly shorter than the predicted value.  On the other hand, the second half period, when it is on average above it, is slightly longer by a few percent.  These differences arise due to the vertical density gradient in the background medium as the cloud bobs up and down, which results in corresponding changes in the B-V times.  Regardless of these minor details, this result confirms the prediction by Binney et al. (2009) in the linear case (or alternately it can be viewed as a test of the simulation method).  When the cooling time is significantly longer than the B-V time, as in this model ({\it h8o1.01}), the cloud does not condense or cool but instead simply oscillates in the vertical direction due to the interplay of gravity and buoyancy.  The velocity steadily damps over the five oscillation periods shown as the cloud deforms and mixes with the background medium.

\subsection{Nonlinear regime}

The results in Binney et al. (2009) were restricted to linear analysis and hence only to the onset of the instability for very low amplitude perturbations.  However, recent studies have suggested that nonlinear perturbations may arise as a natural consequence of the galaxy formation process.  For example, in the cosmological simulation of a Milky Way-like halo by Kere\v{s} \& Hernquist (2009), inflowing filamentary gas cools and forms clouds at distances of 10-100 kpc from the galaxy.  They suggested this as an important fueling mechanism for continued star formation in the disk.  It is therefore interesting to examine how mildly or highly nonlinear clouds evolve over time.

Figure \ref{seq_cool} displays the specific entropy distribution in two of our models with initial cloud heights of 8 kpc.  Each column corresponds to a time slice at {\it (from left to right)} $t = 0$, 18, 25, 35, and 65 Myr.  With an initial overdensity of 4, the cloud does not have sufficient time to cool before being disrupted via the Kelvin-Helmholtz instability (KHI) {\it (left panels)}.  On the other hand, with an initial overdensity of 10 {\it(right panels)}, the cloud does cool and lowers its specific entropy, as can be seen in the third column on the right, at $t = 25$ Myr.  Afterward, again due to the KHI, the cloud mixes with the surrounding hotter gas (last column); however, in this paper we focus on the early evolution of the clouds (i.e. whether they cool or not) and leave a discussion of their ultimate fate to future work.  See also Heitsch \& Putman (2009) for a discussion of the evolution of clouds that are fully formed.

The result shown in Figure~\ref{seq_cool} implies that there is a threshold overdensity, below which the cloud does not cool before disruption and above which it does.  Linear theory does not hold when the cloud overdensity is large, as in these cases; however, we can still develop an analytic expectation based on time-scale arguments.  In the analytic case, the criteria for instability can be phrased in terms of the ratio of the B-V time to the cooling time.\footnote{More precisely, the thermal instability time, although this is generally close to the cooling time (see Binney et al. 2009).}  When $t_{\rm BV}/t_{\rm cool} < 1$, the cloud oscillates rather than cooling.  In the nonlinear case, the gravitational force on the cloud dominates the buoyancy force, and so we neglect it.\footnote{Buoyancy will lead to a timescale correction of order $\delta^{-1}$, where $\delta$ is the cloud overdensity.}  The cloud accelerates relative to the background gas and the resulting velocity difference leads to surface shear and growth of the KHI.  To simplify, we assume the KHI results in the rapid destruction of the clouds once they reach a significant fraction of the sound speed in the background medium, which is ultimately related to the depth of the potential well.  Therefore, we can define an approximate acceleration timescale as
\begin{equation}
t_{\rm accel} = \frac{c_s}{g} = \left(\frac{\gamma k_B T}{ \mu m_p g^2} \right)^{1/2}.
\end{equation}
Notice that for an isothermal halo, this timescale is directly proportional to the B-V time, but is shorter by a factor of 7.7 for $\gamma = 5/3$.  The criteria for nonlinear cloud disruption then becomes approximately $t_{\rm cool}/t_{\rm accel} > 1$.  

This simple calculation neglects a number of important effects, such as the time for cloud disruption (which we will discuss later), as well as heating due to adiabatic compression while the cloud falls.  This last effect can play a role -- for example, Figure \ref{tevol} shows  the time evolution of the maximum density, minimum temperature, and minimum specific entropy in the two models presented in Fig. \ref{seq_cool}.  Unlike in the $\delta=10$ model {\it (right)}, the $\delta=4$ model {\it (left)} does not cool significantly before disruption at $t \approx 60$ Myr.  The density increase due to compression leads to adiabatic heating of the cloud (note that the specific entropy is nearly constant in time in the left column), which prevents cooling.  

\subsubsection{Dependence on cloud height, overdensity, and metallicity}

As we have discussed, the key parameter determining whether cloud disruption occurs (for nonlinear clouds) is the ratio of the cooling time to acceleration time.  We test this idea using a wide range of simulations, as described in Table~\ref{tbl_models}, and shown graphically in Figure \ref{time_ratio}.  The horizontal axis of this figure denotes the overdensity of the initially spherical density perturbation (cloud) relative to the background medium, while the left and right vertical axis show the ratios $t_{\rm cool}/t_{\rm BV}$ and $t_{\rm cool}/t_{\rm accel}$, respectively.  Note that the cloud is set initially in pressure equilibrium with the surrounding gas and hence the temperature inside the cloud is lower by the same overdensity factor -- we use the cloud density and temperature to compute the cooling time, while the background temperature is used in $t_{\rm BV}$ and $t_{\rm accel}$.

Each of the models that we have run are plotted on Figure~\ref{time_ratio}, with the symbol indicating the result of the simulation.  To determine if significant cooling occurs, we look at the minimum specific entropy in the cloud, and determine the ratio of that minimum entropy to the initial entropy.  A cross indicates that no significant cooling occurred (entropy drop less than 20\%), a circle is used for clouds which clearly cooled (entropy drop larger than 50\%), while a triangle is used for intermediate cases (entropy decrease between 20\% and 50\%).  

The results define a clear separation between clouds that can cool and those that cannot, based on the ratio $t_{\rm cool}/t_{\rm accel} \approx 1.0$ (or equivalently, $t_{\rm cool}/t_{\rm BV} = 0.12)$.  For ratios lower than this critical value, the cloud can cool (at least somewhat); while above it, it is disrupted before significant cooling can occur.  This holds true for a wide variety of cloud overdensities and initial heights, and is the primary result of this paper.  

We have assumed the gas to have a metallicity of $0.3 Z_{\odot}$, but the metallicity of the hot halo, or inflowing clouds, is actually uncertain.  High metallicity gas will have shorter cooling times and hence be more likely to condense out than low metallicity gas, while low metallicity has the opposite effect.  To explore this, we have repeated all runs with primordial metallicity (shown as black symbols in Figure~\ref{time_ratio}), and in addition have carried out a run with $Z = 0.1 Z_{\odot}$ and one with $Z = 0.5 Z_{\odot}$ (see Table 1 for results).  The change in metallicity affects the cooling rate but we see that these models are still consistent with the relation defined above (i.e. the ratio of $t_{\rm cool}/t_{\rm accel}$).  

We have also explored the use of the ratio $t_{\rm cool}/t_{\rm dyn}$, employing $t_{\rm dyn} = (2h/g)^{1/2}$ instead of $t_{\rm accel}$ as the disruption timescale.  This ratio is quite good in predicting whether a given cloud will be able to cool before disruption, but does not seem to be as sensitive as $t_{\rm cool}/t_{\rm accel}$ for border-line cases.  This may not be surprising, as the dynamical time is a measure of the time to reach the midplane, while the cloud is sensitive to disruption due to large velocities.

\subsection{Cloud mass}

The KHI timescale ($t_{\rm KH}$) is linearly proportional to the physical size of the perturbing object (e.g., Nulsen 1982).  The {\it disruption} time is usually found to be a few times longer than $t_{\rm KH}$, as the instability develops first on the smallest scales and must grow to a wavelength comparable to the size of the object before it can fully disrupt the cloud.  Therefore, larger clouds might be expected to be able to cool before being disrupted.

To investigate the dependence of cloud disruption time on cloud mass, we ran an additional model with $h_{init} = 32$ kpc ({\it h32o30}) but with twice the usual cloud radius (2 kpc instead of 1 kpc).  We also double the width of the computational volume from $4 \times 4 \times 40$ kpc$^3$ to $8 \times 8 \times 40$ kpc$^3$  ({\it L8h32o30};  see Table~\ref{tbl_models} for exact parameters).  

In comparing the results of the two runs, which differ only in cloud size, we find that the two clouds are accelerated at the same rate, but that the larger cloud takes somewhat longer to disrupt.  This gives more time for cooling to act, and does result in lower entropy and larger (by a factor of $\gtrsim$1.5) travel distance. 
This indicates that for clouds larger than presented here, $t_{\rm cool}/t_{\rm accel}$ may not be the only criteria we need to consider when determining if a cloud will cool or disrupt.
 
 The disruption time of the clouds once they cool and begin moving through the hot halo is also dependent on the mass of the cloud.
Heitsch \& Putman (2009) present a detailed analysis of the disruption time and distance traveled for halo clouds similar to the high velocity clouds at $<10^4$ K that our cooling clouds evolve towards.  In particular, they found that only clouds with masses larger than approximately $10^{4.5}$ \msun\ could make it to the midplane from 10 kpc without being disrupted (see figure 7 of that paper).  In runs where we follow the evolution of the cloud after initially cooling, we also find the condensations do not travel more than 10 kpc before disrupting.  This distance is smaller as the mass of the cloud decreases -- for example, in the {\it h8o10} run shown in Figure~\ref{seq_cool} (with a cloud mass of $1.22 \times 10^3$ M$_{\odot}$), disruption occurs after about 3-4 kpc.

\subsection{Resolution issues}

Cloud disruption by shear instabilities requires high resolution to follow accurately; Klein et al. (1994) showed that in the case of a strong shock sweeping over a cloud, some aspects of the evolution were only resolved with 100 or more cells across the cloud radius, although the disruption timescale could be determined with less resolution.  Our canonical resolution is 64 cells across the cloud radius (we denote this as R64).  In order to investigate the impact of resolution, we have also carried out a simulation of the {\it h32o100} case with R128 (i.e. twice the spatial resolution).  The results are shown in Figure~\ref{resolution} and are very similar for the two different resolution, indicating that we resolve the cooling and disruption.  We chose this run for a resolution study because it has the largest cell size (since the box is the largest), and hence the lowest resolution of any of our standard runs.

\section{Discussion}

What are the implications of these idealized numerical models on fueling of the gaseous galactic disks?
We differentiate between two sources of cold gas accreting on to the disk: (i) gas cooling from the hot halo ($T \sim 10^6$ K) via a linear thermal instability, and (ii) gas cooling from filamentary flows that have intermediate temperatures ($T \sim 10^5$ K) and nonlinear overdensities compared to the hot halo gas they are flowing through.  In section~\ref{sec:linear}, we examine the first idea, as it seems to have been the motivating mechanism behind the Maller \& Bullock (2004) method of solving the over-cooling problem (although it may not be the only way to achieve multi-phase gas).  This mechanism is also explicitly examined in the SPH simulations of Kaufmann \etal (2006) and Sommer-Larsen (2006).
The second source (ii) is discussed in \S4.2 and is relevant to the accretion of cold gas from the IGM or satellites.

\subsection{Implications for classical (linear) hot halo cooling}
\label{sec:linear}

We find that linear perturbations are generally stable in hot galactic halos, in agreement with Binney \etal (2009).  Therefore, unless the halo or cloud properties are quite different from what we model, we do not expect cold gas to form via this mechanism.  
Note that we have only examined local cooling instabilities;  see Walters \& Cox (2001) for the possibility of global vertical oscillations.

Before turning to nonlinear instabilities, we examine how different conditions could change our results.  
First, as discussed in more detail in Binney \etal (2009), it is, in theory, possible to generate linear instabilities via extremely oblate perturbations, which increase the perturbation's inertia and slow the oscillation.  This particular perturbation configuration seems very unlikely and we do not consider it further here.  Another possibility is if the hot halo has a very flat entropy gradient the buoyancy is reduced and this puts the ratio $t_{\rm cool}/t_{BV}$ below 1.  Kaufmann \etal (2009) examined this scenario with an isolated galaxy formation simulation and initialized the gas to have a high uniform entropy, finding that the resulting entropy was very flat and that cool clumps did form (while in a second simulation which had lower initial entropy, they did not).  Such high entropy requires a substantial amount of non-gravitational heating, more than the current generation of cosmological galaxy formation simulations produce (e.g. Kere\v{s} \etal 2009).  Finally, there are a number of physical effects that we did not include in this simulation, including rotation of the halo gas, magnetic fields, and thermal conduction.  We discuss these further in \S~\ref{sec:caveats}, but of all of these, it appears that rotation might be the best chance to rescue linear perturbations, since some degree of rotation support will reduce the buoyancy force and thus the length the B-V timescale.
However, for this to be important, the rotational velocity has to be quite large, probably close to Keplerian.

Since (subject to the above caveats) we do not find linear perturbations to condense out of hot galactic gas, we turn to the question of why the simulations of Kaufmann \etal (2006) and Sommer-Larsen (2006) did.  
The entropy profiles of these SPH calculations are not given, so it is difficult to know if they are significantly flatter than we adopt here, but cosmological collapse simulations generally produce reasonably steep entropy profiles.  The Sommer-Larsen simulation was cosmological, and so did include filamentary flows, and so the perturbations might be seeded by these flows (we address this possibility in more detail in the next section); however, both sets of authors argue that the condensation is coming from the hot gas, and Kaufmann \etal (2006) in particular suggest that the seeds are relatively small fluctuations arising from particle noise.  Based on the simulations here, we do not expect such fluctuations to cool.  One other possible reason for this discrepancy is resolution -- the simulations in this paper have significantly better resolution (several pc) than that in the cosmological runs (typically several hundred pc).  Another possible explanation is the difficulty that SPH codes have resolving Kelvin Helmholtz instabiltiies -- Agertz \etal (2007) found that dense clumps moving through a diffuse background were artificially stabilized by particle effects at the boundary.  This presumably could help explain why slightly overdense clouds were not disrupted, and may act to suppress the oscillations that stabilize linear overdensities.  

\subsection{Implications for cloud formation from (nonlinear) filamentary flows}

We turn now to the evolution of nonlinear clouds -- initial perturbations which are well out of the linear regime.  These might be due to particularly large perturbations from dark matter halos, or, more likely, from inflowing filamentary gas.  In particular, the low-temperature end of the Warm Hot Ionized Medium at $\sim10^5$ K (WHIM; e.g. Dav{\'e} et al. 2001) will have overdensities of order 5-30 when flowing into the hot halo.  This material is sufficiently overdense that it escapes significant heating by the virial accretion shock.  Clumps of this gas might then cool and accrete onto the disk, providing fuel for star formation.  

\begin{figure}
\epsscale{0.88}
\hspace{0mm} \plotone{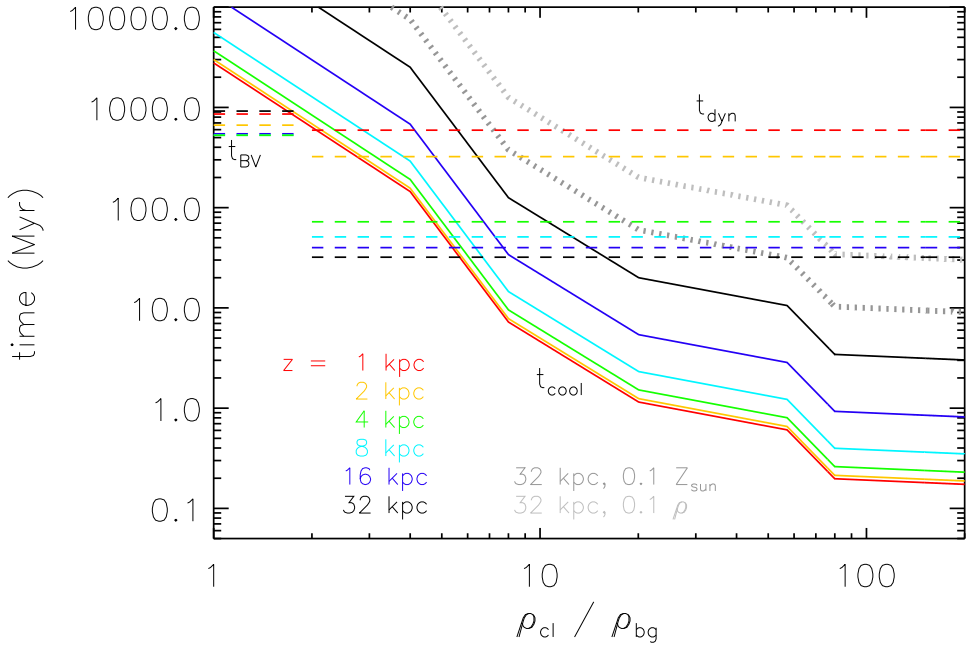}
\caption{The solid lines show cooling times as a function of the initial overdensity for a range of initial cloud heights from 1 to 32 kpc, from bottom to top.  For these curves, we assume a metallicity of 0.3 solar; the thick dotted line in dark grey shows the cooling times for 0.1 solar metallicity gas at 32 kpc, while the other thick dotted line in light grey shows those for a background medium that is only 10\% as dense as in the fiducial model.  Also shown, as dashed lines, are the B-V {\it (left)} and dynamical times {\it (right)} for the same region of initial cloud heights. The exact shape of the cooling times are dictated by the cooling function, $\Lambda(T)$, as the initial temperature is inversely proportional to the overdensity factor (note that pressure equilibrium was assumed).  Hence the cooling times have their minima at $\rho_{cl}/\rho_{bg} \approx 200$, which corresponds to an initial cloud temperature of $T_{\rm halo} \, / \, 200 \approx 10^4$ K.  The B-V and dynamical times do not depend on the cloud overdensity as they are determined by properties of the background medium and the gravitational acceleration at a given height.  The cloud will cool if $t_{cool} < t_{BV}$, otherwise it will oscillate vertically. 
\label{timescale}}
\end{figure}

\begin{figure}
\epsscale{0.78}
\hspace{3mm} \plotone{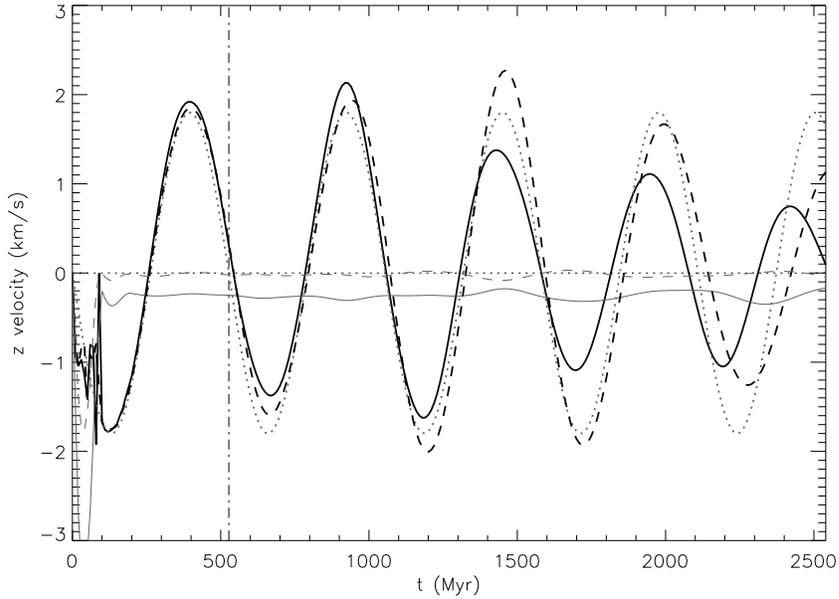}
\caption{The vertical velocity of the cloud as a function of time in model {\it h8o1.01} with $\delta = 0.01$.  The grey dotted curve represents a perfect sinusoidal save with a period of 527.4 Myr (represented by the vertical dot-dashed line) as predicted by linear analysis.  The dark dashed curve shows the mean vertical velocity of the cloud in the model without radiative cooling.  The dark solid curve shows the same quantity in the model with cooling.  In the model with cooling, the cloud slowly sinks due to cooling of the atmosphere below the cloud.  To make the plot clearer, we have removed this nearly constant negative z-velocity from the plotted curves.  It was measured by averaging the vertical velocity well above and below the cloud, and is shown as the light grey solid and dashed lines for the cooling and no-cooling cases, respectively.
\label{oscillate}}
\end{figure}

\begin{figure}
\epsscale{1.0}
\hspace{0mm} \plotone{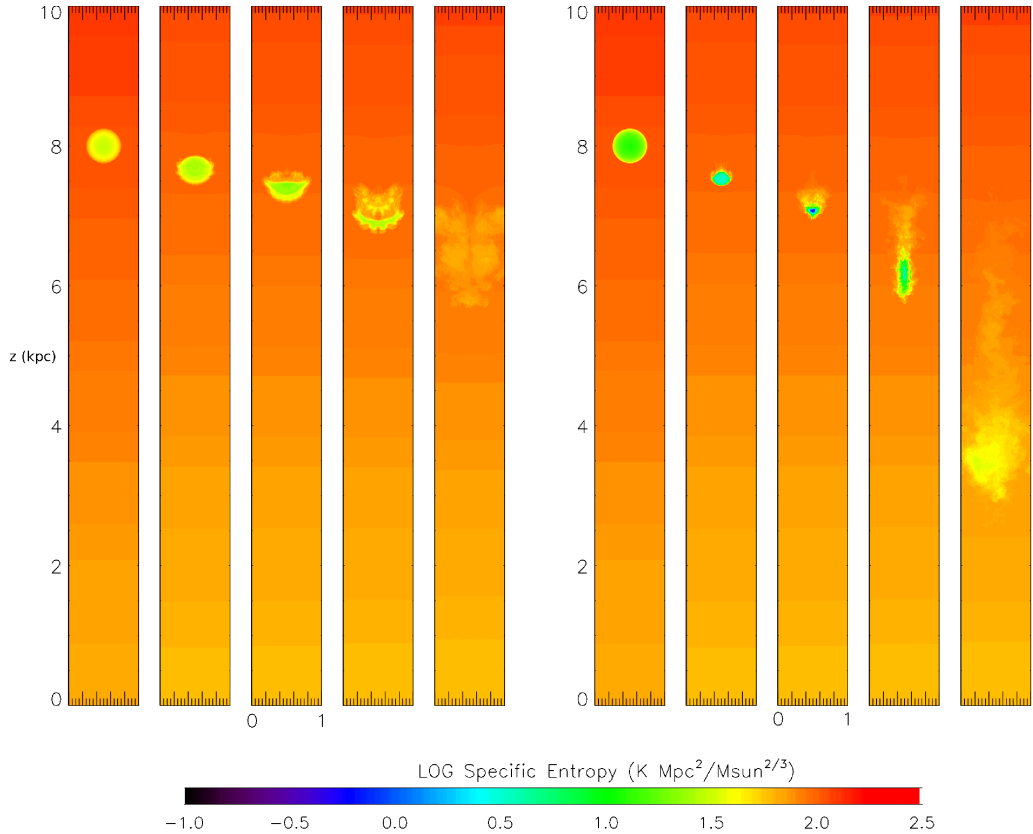}
\caption{Specific entropy distribution in two of our models with the initial cloud height of 8 kpc.  Each column corresponds to a time slice at {\it (from left to right)} $t = 0$, 18, 25, 35, and 65 Myr.  (a) With an initial overdensity of 4, the cloud does not have sufficient time to cool before being disrupted via KHI.  (b) With an initial overdensity of 10, the cloud cools and condenses, lowering the specific entropy of the cloud (see the third column).  Afterward, again due to KHI, the cloud mixes with the surrounding hotter gas (last column).  The colorbar shown at the bottom applies to both models.  
\label{seq_cool}}
\end{figure}

\begin{figure}
\epsscale{1.0}
\hspace{0mm} \plotone{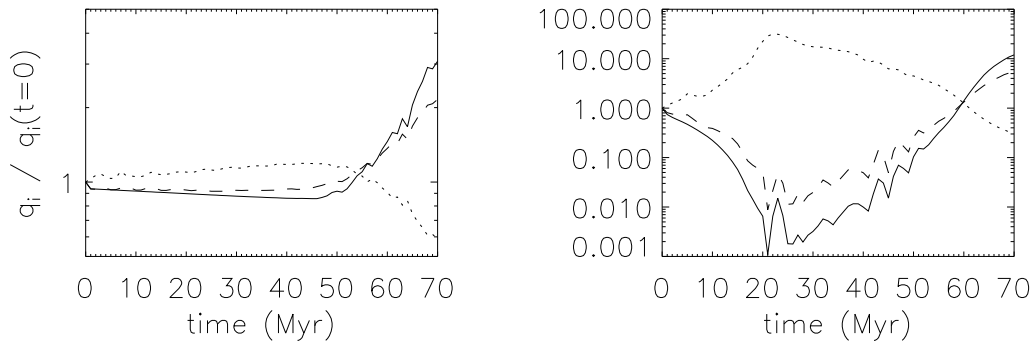}
\caption{Time evolution of maximum density (dotted), minimum temperature (dashed), and minimum specific entropy (solid) in the two models presented in Fig. \ref{seq_cool}, showing the fractional change from their respective initial values.  (Note that the vertical ranges are different in the two plots.)  Unlike in the $\delta=10$ model {\it (right)}, the $\delta=4$ model {\it (left)} does not cool significantly before disruption at $t \approx 60$ Myr.  The density increase due to compression leads to adiabatic heating of the cloud (note that the specific entropy is nearly constant in time in the left column), which slows cooling.  
For these figures, cells with the minimum 3\% in terms of specific entropy were used, with averages weighted by mass.
\label{tevol}}
\end{figure}

\begin{figure}
\epsscale{1.0}
\hspace{0mm} \plotone{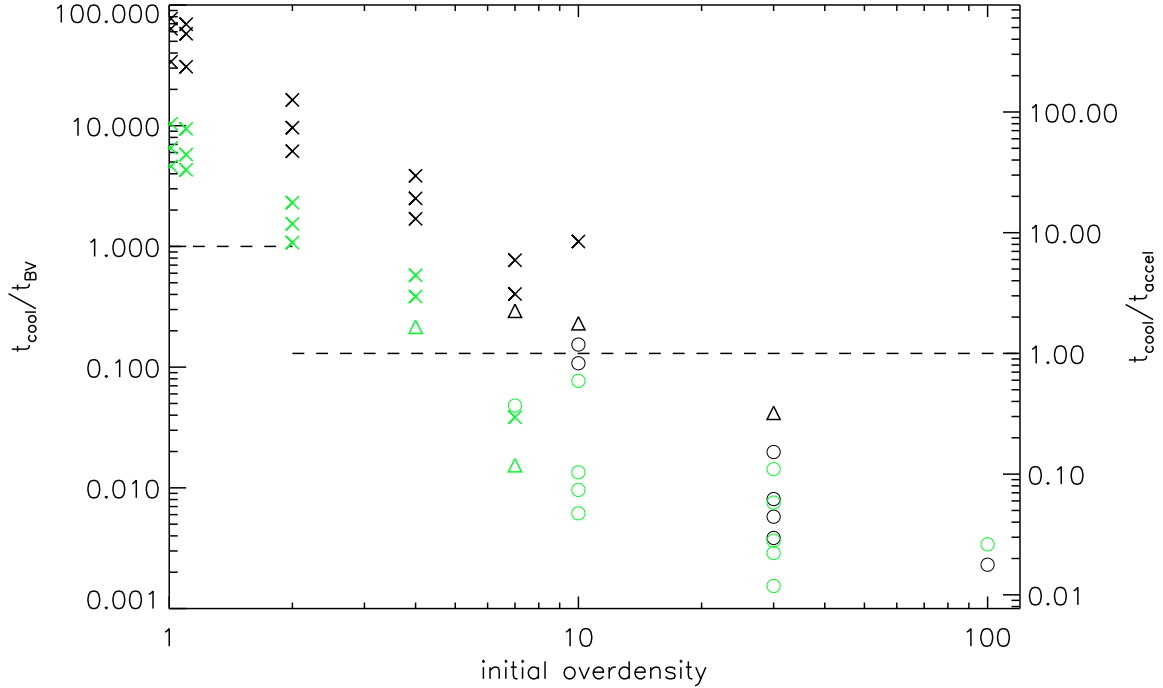}
\caption{We plot the models we have run in the space of overdensity and the ratio of the cooling time to the B-V time {\it (left)} scale, or to $t_{\rm accel}$ {\it (right)} scale.  Horizontal dashed lines show unity ratios for both scales.  Crosses indicates runs that clearly did not cool (entropy drop by less than 20\%), triangles are marginal cases (entropy drop between 20\% and 50\%, while circles are runs which show clear cooling (central entropy drop larger than 50\%).  Black symbols assume primordial abundances, while green symbols are used for metallicity of $Z = 0.3 Z_{\odot}$.  
}
\label{time_ratio}
\end{figure}

\begin{figure}
\epsscale{1.0}
\hspace{0mm} \plotone{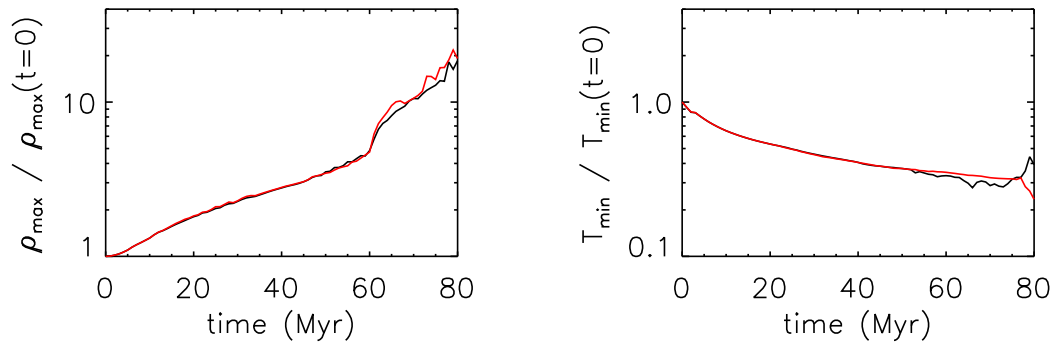}
\caption{A comparison of the maximum density and minimum temperature for the {\it h32o100} run at two different resolutions: 64 cells across the cloud radius (R64, black) and 128 cells across the cloud radius (R128, red).  Noticeable differences are seen only after $t \approx 60$ Myr, when the cloud is gradually disrupted due to the KHI, which is not the main focus of this paper.}
\label{resolution}
\end{figure}

We find that overdensities can cool significantly, provided they meet the criteria that the ratio of the cooling to acceleration time is below about 1 (see Fig.~\ref{time_ratio}).  This generally entails substantial overdensities and/or low initial cloud heights (since the cooling time drops faster than the acceleration time with decreasing height).  For accretion of WHIM-like gas in filaments at distances around 40 kpc, as suggested by Kere\v{s} \& Hernquist (2009), Figures~\ref{timescale} and \ref{time_ratio} indicate that overdensities of about 10 are required.  The overdensities in the filament in that paper appear to be similar to this value, therefore we conclude that this mechanism for nonlinear seeding of cooling instabilities may be a viable way to produce cold halo gas.  This is one of the key results of this paper.  We also note that we monitored the material cooling and find that the surrounding hot halo medium does not cool along with the overdensity.  

The clouds modeled here are initially most similar in temperature to the O{\small~VI} absorbers detected with ultraviolet spectrographs ($\sim$10$^5$ K; Thom \& Chen 2008; Tripp et al. 2008; Prochaska et al. 2011; Narayanan et al. 2011).  The sample of O{\small~VI} absorbers  
remains somewhat limited, with detections thus far largely beyond 100 kpc (in projection) from the nearest galaxy.  For most systems it  
remains somewhat unclear if the absorbers are arising from collisional or photo-ionization.   Recently there has been a detection of an O{\small~VI} absorber at $z=0.35$ only 85 km/s and 95 kpc in projection from a galaxy that is found to be a photoionized structure with a length scale $L \approx$ 0.1--1.2 Mpc and log $n_{\rm H} = -4.4$ to $-4.9$ (Thom et al. 2011).   This density is typical of the densities estimated for other potentially photoionized O{\small~VI} absorbers (log $n_{\rm H} = -4$ to $< -5$; Thom \& Chen 2008; Prochaska et al. 2011).   Based on our results, these types of low density structures may have difficulty being able to cool.  For $\sim$10$^5$ K flows from the IGM to play a substantial role in feeding galaxy halos cold gas, we would expect the future absorbers detected by COS to be found closer to galaxies and with somewhat higher overdensities.  Our results may also have implications on the cooling of warmer clouds traced by Ly$\alpha$ and Mg{\small~II} absorption line systems, but further simulations with initial cloud conditions more similar to these should be completed before definitive statements can be made.

For those clouds that do cool before disrupting, they usually reach a minimum temperature around $10^4$ K (although in some cases $T_{min} < 10^3$ K) and continue to accelerate.  They eventually reach large velocities, and are subject to dynamical instabilities which, in our simulations, quickly shred the clouds.  This can be seen in the last panel in Fig.~\ref{seq_cool}, where the cloud has been dispersed and the averaged specific entropy starts to grow with time.  This occurs after the cloud has travelled only about 10 kpc, and so occurs before it hits the disk.  This is in agreement with Heitsch \& Putman (2009), who also find (for similar cloud masses), a maximum distance before disruption of about 10 kpc.  This seems to imply that even if the cloud cools, it will not be able to fuel the disk with cold gas, and will instead simply add to the hot halo.

The cloud mass determines whether or not the cloud will be disrupted before it reaches the midplane, which takes roughly a dynamical time.  
The larger the mass, the longer the disruption length, and so we suspect that clouds with even larger masses (than the maximum value of $2.07 \times 10^5$ M$_\odot$ in our models) will travel further, although this has not specifically been tested.   In Kere\v{s} \& Hernquist, the typical cloud masses were about $10^6$ M$_\odot$, however this may be largely due to their resolution limitations.  In any case, it would be interesting to further investigate the survival of such large clouds in future work as our results indicate the direct accretion of gas to the disk from cold flows or stripped satellite gas at large radii may be difficult despite the initial ability for clumps to cool. 

Finally, we note that there is an interesting regime where the cooling time is shorter than the B-V or the acceleration time but longer than the dynamical time.  In this case, the gas will start to cool but will impact the disk before it can cool significantly, and so will be accreted as warm gas.  Such inflowing gas will escape H{\small~I} observations and instead appear as an ionized component with negative velocities (e.g., Haffner et al. 2003).  For an initial height of 32 kpc ({\it h32}) and 0.3 solar metallicity gas, this is the case for clouds with initial overdensities between $\sim$8 and $\sim$15 (see Fig. \ref{timescale};  for primordial gas, the range is between $\sim$20 and $\sim$40).  Two of our models reside in this regime ({\it h32o30} and {\it L8h32o30} with $Z=0$);  both models indeed show that the clouds stay warm, but only in the latter model is it not fully disrupted until impacting the midplane.

\subsection{Physical Processes Not Included In the Simulations}
\label{sec:caveats}

There are a number of physical effects that we do not include in our models -- we discuss their possible impact in this section, starting with magnetic fields, then heat conduction, and rotation.

The impact of the magnetic field on the linear instability is subtle -- Loewenstein (1990) and Balbus (1991) show that, for radial fields, the field can suppress the buoyancy oscillations and so enhance the cooling instability.  However, purely radial fields do not seem a natural outcome since they are dynamically weak and even mild turbulence will tangle them, leading only to a small contribution to the effective isotropic pressure, and negligible effect on the instability.

Magnetic fields may act to stabilize a nonlinear perturbation against disruption.   For the nonlinear case, fields may insulate the cloud (e.g. Konz \etal 2002; Dursi \& Pfrommer 2008), although Stone \& Gardiner (2007) did not find that a dynamically weak field significantly suppressed the growth of Rayleigh-Taylor (R-T) instabilities.  Recently, McClure-Griffiths et al. (2010) measured a parallel component of the magnetic field inside an HVC associated with the Magellanic System to be greater than $\sim$6 $\mu$G, which they argue is more than sufficient to stabilize it against ram pressure stripping.  Fields of similar strength were not found for 26 other HVCs in this work, but this may be partially due to complicated foregrounds.   It will be an interesting future study to explore similar models in a magnetized medium.   

In addition, thermal conduction, which we do not include, can act to suppress the cooling of clouds.  This is treated for linear perturbations in Binney et al (2009), but an investigation for nonlinear clouds would also be interesting.  We expect this would make the clouds even more susceptible to disruption, making our results for the minimum overdensity for cooling a lower limit.

Rotation can also impact the stability of linear clouds, as centrifugal support will decrease the level of pressure support required and therefore lengthen the duration of buoyancy oscillations.  The level of rotation required for this to be important is probably quite large, and so may play a role close to the disk, but will be less important at large radii.

Finally, we note that the clouds we model have been spherical with a limited range of sizes, and zero initial velocity.  A more complete investigation of the impact of cloud size, initial velocity and cloud shape would be useful, but beyond the scope of this paper.   

\section{Conclusions}

Using AMR hydrodynamic simulations of vertically stratified hot halo gas, we examine the conditions under which clouds can form and condense out to potentially fuel star formation in the gaseous disk.

\begin{itemize}

\item We find that halo clouds do not develop from linear perturbations in density and temperature.  This is a regime where the cooling time is longer than the B-V time and hence predicted to be stable by Binney et al. (2009).  Although we did not consider this case, if the cooling time were shorter than the B-V time at a given height, the entire layer -- as opposed to small condensed clouds in the background medium -- would have cooled and fallen down to the midplane, resulting in a global catastrophic collapse.

\item We extend the theory into the nonlinear regime by considering mildly or strongly nonlinear perturbations with overdensities up to 30 (100 in one model with $h_{\rm init} = 32$ kpc).  Here, the result depends on the ratio of cooling time to the time it takes to accelerate the cloud to a significant fraction of the sound speed (i.e. $t_{\rm cool}/t_{\rm accel}$).  If this ratio is below a critical value, around 1, the cloud can cool significantly before it is disrupted by K-H and/or R-T instabilities and may reach high enough column densities to be detectable in HI surveys ($N_{HI} \gtrsim 10^{18.5}$ cm$^{-2}$).  

\item We simulate a wide range of cloud overdensities and initial heights, and show that the $t_{\rm cool}/t_{\rm accel}$ criteria is a good predictor of whether or not a seeded cloud will be able to cool significantly.  We show this result scales as expected for different metallicities, and is a numerically well-resolved result.  We also investigate cloud size and find that, once formed, large clouds take longer to disrupt.  


\item Nonlinear perturbations, such as those seeded by filamentary flows from cosmological simulations (e.g. Kere\v{s} and Hernquist 2009), appear to be right at the critical threshold and therefore probably can cool before being disrupted.  However, we find (in agreement with Heitsch \& Putman (2009)) that clouds with $M \lesssim 10^5$ M$_\odot$ are disrupted within 10 kpc, and suspect that only larger mass clouds can potentially survive the trip to the disk.  


\end{itemize}


\acknowledgements

We acknowledge support from NSF grants AST-05-47823, AST-09-08390, AST-09-04059, and AST-10-08134, as well as computational resources from NASAÕs Pleiades, which is provided by the NASA High-End Computing (HEC) Program through the NASA Advanced Supercomputing (NAS) Division at Ames Research Center, NSF Teragrid, and Columbia University's Hotfoot cluster.  We thank Josh Peek, Fabian Heitsch, Mordecai-Mark Mac Low, and Dusan Kere\v{s} for useful discussions.



\clearpage

\begin{deluxetable}{c|c|c|c|c|c|c|c}
\tablecaption{Parameters for the simulations\label{tbl_models}}
\tablewidth{0pt}
\tablehead{
\colhead{model}  & \colhead{$h_{\rm init}$ (kpc)} & \colhead{$n_{\rm bg}$
$(cm^{-3})$} & \colhead{$\delta$} & \colhead{$Z$ ($Z_{\odot}$)} &
\colhead{$R_{\rm cloud}$ (kpc)} & \colhead{$M_{\rm cloud}$ (M$_{\odot}$)} & \colhead{Cooled?}
}
\startdata
$h2o1.01$        &   2 & $2.35 \times 10^{-4}$ &   1.01 & 0, 0.3 & 0.25 &
$2.34 \times 10^2$ & N, N \\
$h2o1.1$          &       &                                        &   1.1
 &            &        & $2.54 \times 10^2$ & N, N \\
$h2o2$             &       &                                        &     
2    &            &        & $4.62 \times 10^2$ & N, N \\
$h2o4$             &       &                                        &     
4    &            &        & $9.25 \times 10^2$ & N, P \\
$h2o7$             &       &                                        &     
7    &            &        & $1.62 \times 10^3$ & P, P \\
$h2o10$           &       &                                        &    10
  &           &        & $2.31 \times 10^3$ & Y, Y \\
$h2o30$           &       &                                        &    30
  &           &        & $6.94 \times 10^3$ & Y, Y \\
$h4o1.01$        &   4 & $1.91 \times 10^{-4}$ & 1.01   & 0, 0.3 & 0.25 &
$1.90 \times 10^2$ & N, N \\
$h4o1.1$          &      &                                         & 1.1  
  &           &        & $2.07 \times 10^2$ & N, N \\
$h4o2$             &      &                                         &    2
    &            &        & $3.76 \times 10^2$ & N, N \\
$h4o4$             &      &                                         &    4
    &            &        & $7.52 \times 10^2$ & N, N \\
$h4o7$             &      &                                         &    7
    &            &        & $1.32 \times 10^3$ & N, N \\
$h4o10$           &      &                                         &  10  
  &           &        & $1.88 \times 10^3$ & Y, Y \\
$h4o30$           &      &                                         &  30  
  &           &        & $5.64 \times 10^3$ & Y, Y \\
$h8o1.01$        &   8 & $1.24 \times 10^{-4}$ & 1.01   & 0, 0.3 & 0.25 &
$1.23 \times 10^2$ & N, N \\
$h8o1.1$          &       &                                        & 1.1  
 &            &        & $1.34 \times 10^2$ & N, N \\
$h8o2$             &       &                                        &    2
    &            &        & $2.44 \times 10^2$ & N, N \\
$h8o4$             &       &                                        &    4
    &            &        & $4.88 \times 10^2$ & N, N \\
$h8o7$             &       &                                        &    7
    &            &        & $8.54 \times 10^2$ & N, Y \\
$h8o10$           &       &                                        &  10  
 &            &        & $1.22 \times 10^3$ & P, Y \\
$h8o30$           &       &                                        &  30  
 &            &        & $3.66 \times 10^3$ & Y, Y \\
$h16o30$         &  16 & $5.20 \times 10^{-5}$ & 30     &            & 0.5  
& $1.23 \times 10^4$ & P, Y \\
$h32o10$         &  32 & $1.37 \times 10^{-5}$ & 10     &            & 1  
& $8.63 \times 10^3$ & N, Y \\
$h32o30$         &       &                                         & 30   
&            &       & $2.59 \times 10^4$ & P, Y \\
$h32o100$        &      &                                         & 100  &
          &        & $8.63 \times 10^4$ & Y, Y \\
$h32o30z0.1$   &      &                                        &  30    &
0.1     &        & $2.59 \times 10^4$ & Y \\
$h32o30z0.5$   &      &                                        &         
& 0.5      &        &       & Y \\
$h32o30rh$       &      &                                        &        
& 0.3      & 0.5     & $3.24 \times 10^3$ & Y \\
$h32o30rd$       &      &                                        &        
&            & 2      & $2.07 \times 10^5$ & Y \\
$L8h32o30$       &      &                                        &        
&  0, 0.3 &         &       & P, Y \\
\enddata
\tablecomments{The columns show the model name, initial cloud height,
number density of the background medium at $h_{\rm init}$, cloud
overdensity, assumed metallicity of gas, initial cloud radius, and whether
the cloud cooled significantly (Y), partly (P) or not (N).  Blank
indicates the value remains the same as that in the above row.}

\end{deluxetable}

\end{document}